\documentclass[apjl]{emulateapj}
\usepackage{apjfonts}
\usepackage{amsmath, amsthm, amssymb}

\newcommand{\fuvcenter}{1530\AA}
\newcommand{\nuvcenter}{2310\AA}

\newcommand{\fuvmag}{\ifmmode{FUV}\else{\it FUV}}
\newcommand{\nuvmag}{\ifmmode{NUV}\else{\it NUV}}

\newcommand{\etal}{{\sl{}et~al.}}
\newcommand{\gsim}{\rlap{\raise -.3ex\hbox{${\scriptstyle\sim}$}}%
                   \raise .6ex\hbox{${\scriptstyle >}$}}%
\newcommand{\lsim}{\rlap{\raise -.3ex\hbox{${\scriptstyle\sim}$}}%
                   \raise .6ex\hbox{${\scriptstyle <}$}}%

\shortauthors{Budav\'ari et~al.}
\journalinfo{To Appear in the {\sl GALEX} special edition of ApJ Letters}
\submitted{Received: 2004 May 6 \ \ \ \ \ \ Accepted: 2004 June 10}
\begin{document}

\title{The Ultraviolet Luminosity Function of GALEX Galaxies at \\
Photometric Redshifts Between 0.07 and 0.25}
%\shorttitle{UV Luminosity Function of GALEX Galaxies from Photometric Redshifts}

\author{Tam\'as Budav\'ari\altaffilmark{1}, 
Alex S. Szalay\altaffilmark{1},
St\'ephane Charlot\altaffilmark{2,3},
Mark Seibert\altaffilmark{4},
Ted K. Wyder\altaffilmark{4},
St\'ephane Arnouts\altaffilmark{5},
Tom A. Barlow\altaffilmark{4}, 
Luciana Bianchi\altaffilmark{1},
Yong-Ik Byun\altaffilmark{6}, 
Jos\'e Donas\altaffilmark{5},
Karl Forster\altaffilmark{4},
Peter G. Friedman\altaffilmark{4},
Timothy M. Heckman\altaffilmark{1},
Patrick N. Jelinsky\altaffilmark{7},
Young-Wook Lee\altaffilmark{6}, 
Barry F. Madore\altaffilmark{8},
Roger F. Malina\altaffilmark{5},
D. Christopher Martin\altaffilmark{4}, 
Bruno Milliard\altaffilmark{5},
Patrick Morrissey\altaffilmark{4}, 
Susan G. Neff\altaffilmark{9},
R. Michael Rich\altaffilmark{10},
David Schiminovich\altaffilmark{11},
Oswald H. W. Siegmund\altaffilmark{7}, 
Todd Small\altaffilmark{4},
Marie A. Treyer\altaffilmark{4,5},
and
Barry Welsh\altaffilmark{7}
}

\email{budavari@jhu.edu}

\altaffiltext{1}{Department of Physics and Astronomy, The Johns Hopkins
University, 3701 San Martin Drive, Baltimore, MD 21218, USA}

\altaffiltext{2}{Max-Planck-Institute f\"ur Astrophysik,
Karl-Schwarzschild-Strasse 1, 85748 Garching, Germany}

\altaffiltext{3}{Institut d'Astrophysique de Paris, CNRS, 98 bis
boulevard Arago, F-75014 Paris, France}

\altaffiltext{4}{California Institute of Technology, MC 405-47, 1200
E. California Blvd., Pasadena, CA 91125}

\altaffiltext{5}{Laboratoire d'Astrophysique de Marseille, BP 8, Traverse 
du Siphon, 13376 Marseille Cedex 12, France}

\altaffiltext{6}{Center for Space Astrophysics, Yonsei University, Seoul
120-749, Korea}

\altaffiltext{7}{Space Sciences Laboratory, University of California at
Berkeley, 601 Campbell Hall, Berkeley, CA 94720}

\altaffiltext{8}{Observatories of the Carnegie Institution of Washington,
813 Santa Barbara St., Pasadena, CA 91101}

\altaffiltext{9}{Laboratory for Astronomy and Solar Physics, NASA Goddard
Space Flight Center, Greenbelt, MD 20771}

\altaffiltext{10}{Department of Physics and Astronomy, University of
California, Los Angeles, CA 90095}

\altaffiltext{11}{Department of Astronomy, Columbia University, New
York, NY 10027, USA}

\begin{abstract} 
We present measurements of the UV galaxy luminosity function and the
evolution of luminosity density from GALEX observations matched to the
Sloan Digital Sky Survey (SDSS). We analyze galaxies in the Medium
Imaging Survey overlapping the SDSS DR1 with a total coverage of 44
$\deg^2$. Using the combined GALEX+SDSS photometry, we compute
photometric redshifts and study the LF in three redshift shells
between $z=0.07$ and 0.25. The Schechter function fits indicate that
the faint-end slope $\alpha$ is consistent with $-1.1$ at all
redshifts but the characteristic UV luminosity $M^*$ brightens by 0.2
mag from $z=0.07$ to 0.25. In the lowest redshift bin, early and
late type galaxies are studied separately and we confirm that red
galaxies tend to be brighter and have a shallower slope $\alpha$ than
blue ones. The derived luminosity densities are consistent with other
GALEX results based on a local spectroscopic sample from 2dF and the
evolution follows the trend reported by deeper studies.
\end{abstract}

\keywords{ultraviolet: galaxies --- surveys --- galaxies: luminosity
function, evolution}

\section{Introduction} 

In the era of precision cosmology, the star formation history of the
universe can be studied accurately as one can detect evolutionary
effects in the observables on top of the global expansion. In
particular, the restframe ultraviolet luminosity of galaxies has
proven to yield a good handle on the star formation rate
\citep{kennicutt98}. A number of galaxy surveys have probed the
history of star formation at different redshifts. While most studies
agree on a relatively rapid rise in the star formation rate (SFR) out
to redshift of $z \sim 1$, significant uncertainties remain even at
lower redshifts \citep{lilly96,connolly97,cowie99,wilson02}.
The restframe UV continuum of local galaxies is not accessible
from the ground. The balloon-borne telescope of the FOCA experiment
\citep{milliard92} had been the best window onto the UV sky until last
year, when the {\sl Galaxy Evolution Explorer} (GALEX)
satellite was successfully launched to orbit.
This paper is one in the first series of luminosity function papers on
GALEX sources and focuses on galaxies at redshifts between
$z = 0.07$ and $0.25$.  We use photometric redshifts to boost our
sample size by a factor of 20 compared to spectroscopic data
available.  Throughout the paper, we assume a flat $\Lambda$CDM
cosmology with $\Omega_M = 0.3$ and
$H_0 = 70$ km s$^{-1}$ Mpc$^{-1}$.

\section{The Sample}

The GALEX telescope has two photometric bands at \fuvcenter{}
(\fuvmag{}) and \nuvcenter{} (\nuvmag) and a 1.2 degree field of
view. For a detailed description of survey and performance, see
\citet{martin04} and \citet{morrissey04} in the present volume.
Our sample consists of Medium Imaging Survey (MIS) fields overlapping
with the Data Release One coverage of the SDSS \citep{dr1}, see also
\citet{seibert04}. The depth of the MIS fields is well matched to SDSS
and this unique 7-band multicolor dataset provides a good basis for
various statistical studies.  We select the 57 fields with more than
$1400$ second exposure times and no objects with higher extinction
than $E(B\!-\!V)=0.08$. We use 36 arcmin radius circles in the center
of the fields to ensure uniform image quality. The intersection of the
unique area of these MIS fields with the SDSS DR1 footprint is 43.9
square degrees. Our catalog contains only objects that are classified
as galaxies by the SDSS photometric pipeline based on their
morphology. We use total magnitudes corrected for foreground
extinction: SExtractor's {\tt{}MAG\_AUTO} for GALEX and model
magnitudes from SDSS. For the limiting magnitudes, we elect to choose
a safe $m_{\rm{}lim}=21.5$ cut in both bands to ensure completeness
\citep{xu04}.

\begin{figure}[]
\epsscale{1.15}
\plotone{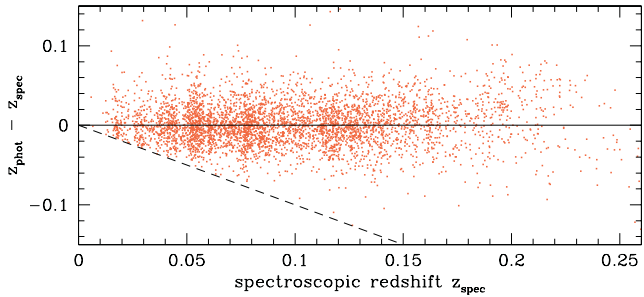}
\caption{Comparison of the spectroscopic and photometric redshifts for
the MIS objects in our training set. The equation of the black dashed
line is $z_{\rm{}phot}=0$.}
\label{fig:zz}
\end{figure}

\subsection{Photometric Redshifts}

Photometric redshifts are utilized to fully exploit the data set. We
choose empirical photometric redshifts over template based estimates
because currently the GALEX photometric system is only known to about
10\% accuracy and SED fitting is sensitive to zeropoint errors.
Following \citet{connolly95a}, a third order polynomial formula was
applied to map the GALEX \nuvmag{} and SDSS $u'g'r'i'z'$ magnitudes to
photometric redshifts. Note that the \fuvmag{} magnitude was excluded
from the fitting formula as the \nuvmag{} observations go deeper and
not all galaxies have \fuvmag{} measurements. This way there is only
one redshift estimator that can be used for both \fuvmag{} and
\nuvmag{} limited samples.  We find that the empirical fit yields
reliable redshift estimates out to redshift of 0.25. For the training
set of 6295 galaxies, the rms scatter is $\Delta z_{\rm{}rms} = 0.026$
and there is about 2\% outliers. This accuracy is about 15\% better
than SDSS alone using the same technique.  A more detailed analysis
and a photometric redshift catalog will be published elsewhere
\citep{budavari04c}.  We expect the uncertainty in the photometric
redshifts to be a significant source of error in our statistical
analysis, so we adopt a conservative nominal redshift error of
$\sigma_z = 0.03$.  Figure~\ref{fig:zz} compares the spectroscopic and
photometric redshifts as a function of redshift.

One of the disadvantages of the empirical photometric redshifts is
that the method does not provide a direct measurement of the spectral
types or K-corrections. To overcome this, we fit synthetic model
spectra with full wavelength coverage from the UV to the IR to the
SDSS photometry and pick the best fitting template for each
galaxy. Our template set has 10 interpolated spectra from Ell to Irr
of \citet{bc03}.  For each of these templates, the K-correction is
calculated as a function of redshift.  In addition to the galaxy
templates, we also include a series of QSO spectra in an attempt to
identify AGNs in the sample. Those objects which are best fitted with
quasar templates, roughly 10\%, are removed from the sample.
Our photometric redshift catalog contains 190,489 MIS galaxies out of
which 9,356 pass the area, magnitude, redshift and SED cuts in the
\nuvmag{} and 6,174 in \fuvmag{}.

\section{Luminosity Function Results}

There are several methods for calculating the luminosity function
\citep{schmidt68,lyndenbell71,choloniewski86,subbarao96}. We use the
$V_{\rm{}max}$ method \citep{schmidt68} to calculate the LF in 0.1
magnitude wide bins, First, we derive the absolute magnitude using the
distance modulus and the K-correction, then the maximum redshift where
the object could be observed from.  The LF is then calculated as
$\phi(M)\,dM = \sum 1 / V(z_{\max})$, where $V(z) = \frac{\Omega}{3}
d^3(z)$ for a flat universe, $\Omega$ is the areal coverage and $d(z)$
is the comoving distance.

To estimate the uncertainty in the LF, we create 50 Monte-Carlo (MC)
realizations of the catalogs drawing the redshifts randomly from
Gaussian distributions with means of the originally estimated
redshifts and widths of $\sigma_z = 0.03$. We coadd the MC
realizations for a more robust estimate of the true LF. This method
allows us to propagate the errors. The errorbars plotted in the
figures are combinations of the variations among the MC realizations
and the Poisson errors added in quadrature.

\subsection{Evolution with Redshift and Spectral Type}

\begin{figure*}[]
\epsscale{1.05}
\plottwo{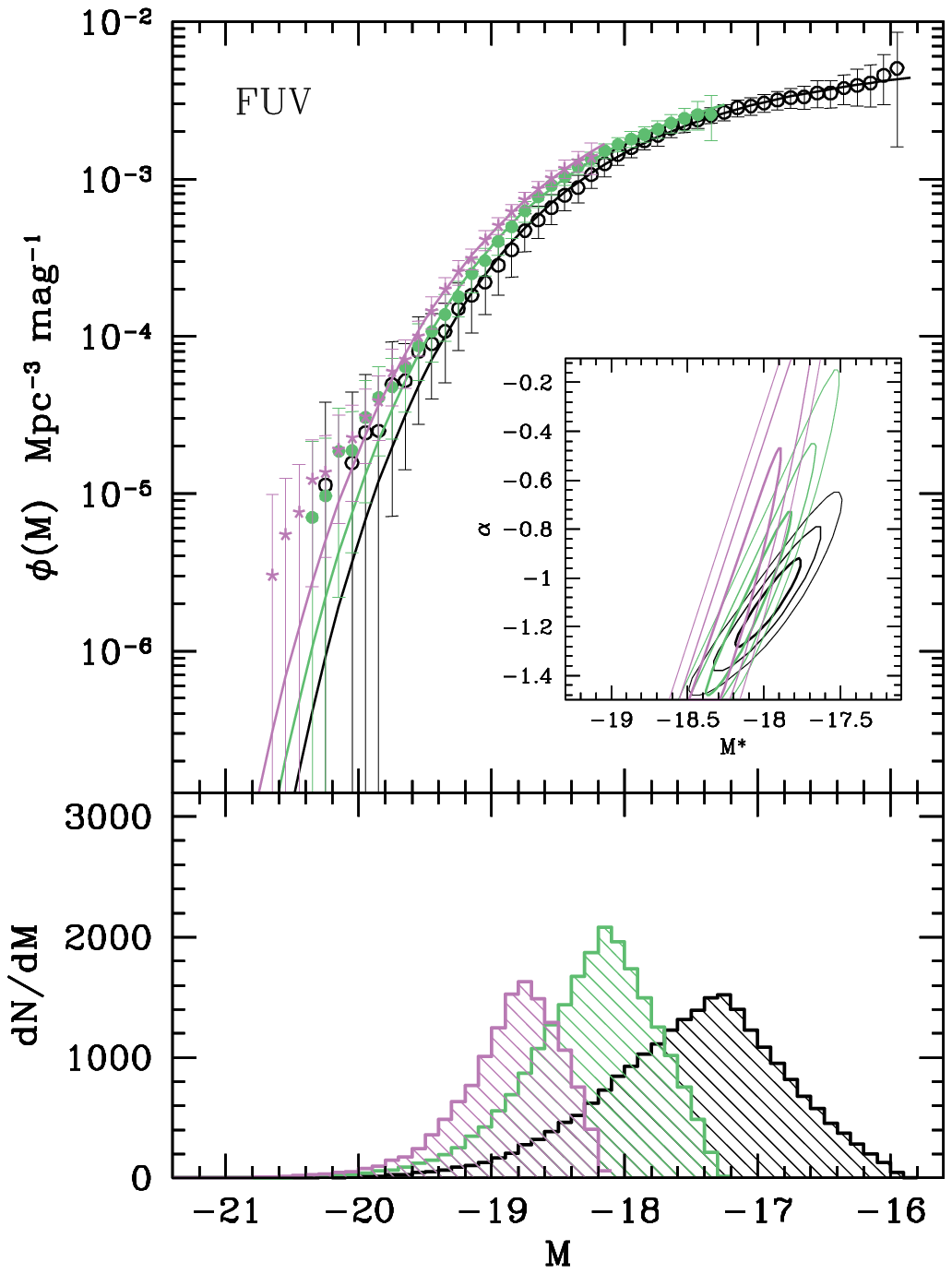}{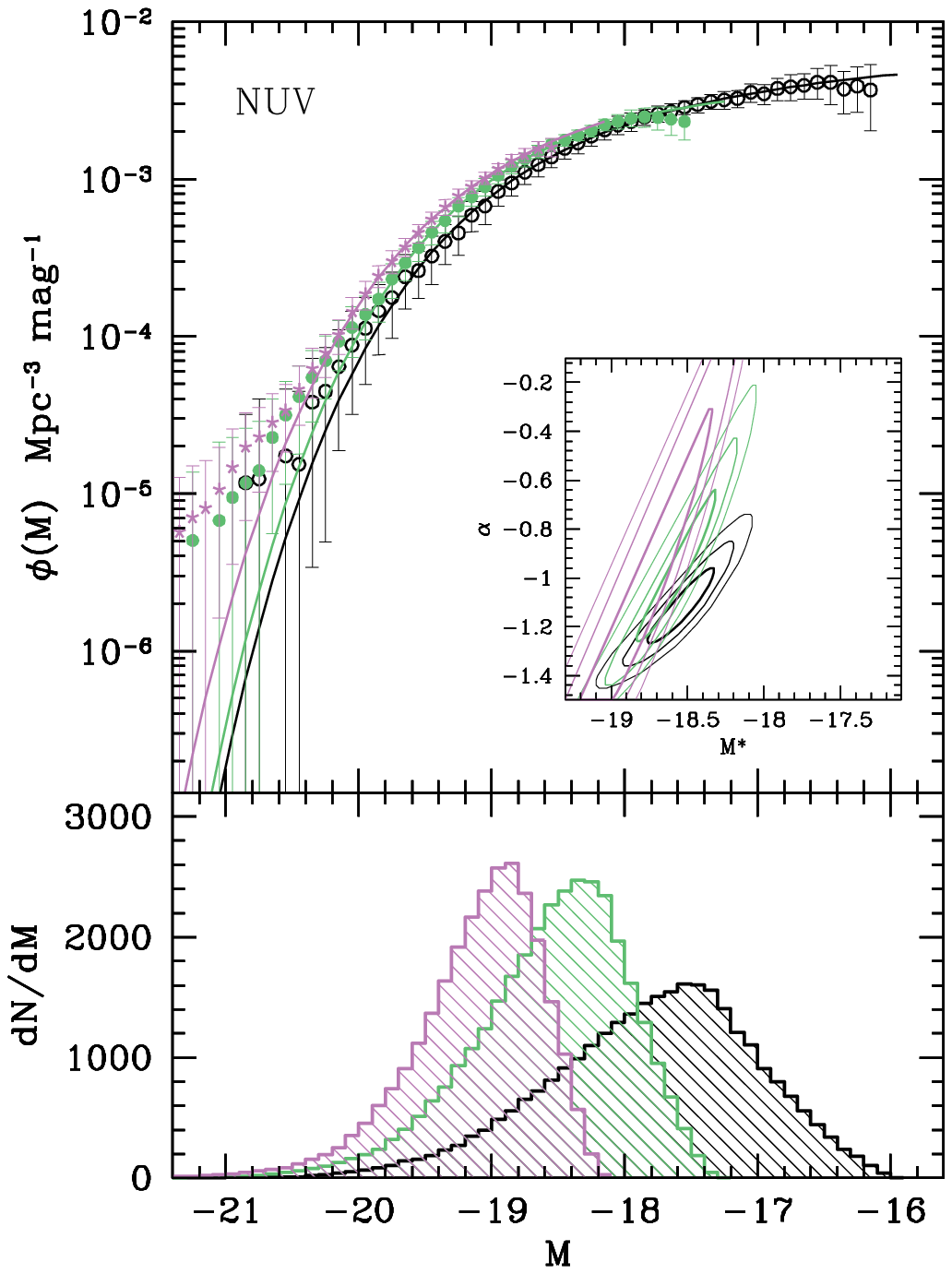}
\caption{The GALEX \fuvmag{} and \nuvmag{} luminosity functions in
three redshift bins ({\it{}black open circle:} $0.07 < z < 0.13$,
{\it{}green solid circle:} $0.13 < z < 0.19$, {\it{}magenta star:}
$0.19 < z < 0.25$.)  The top panel illustrates the $1/V_{\rm{}max}$
measurements along with the best fit Schechter functions. The
confidence regions on $M^*$ and $\alpha$ are shown in the insets. The
bottom panel shows the number of objects involved in the analysis for
a particular selection.}
\label{fig:lfz}
\end{figure*}

To study the evolution of the UV LF as a function of redshift, we
split the sample into three redshift shells. These low, medium and
high redshift subsamples have galaxies in the 0.07--0.13, 0.13--0.19
and 0.19--0.25 intervals.  Figure~\ref{fig:lfz} shows the \fuvmag{}
and \nuvmag{} LFs for the three redshift slices along with their best
fitting \citet{schechter76} functions.  The absolute magnitude range
over the luminosity functions can be fitted is limited at the faint
end by the lower redshift cutoff in the more distance shells and also
at the bright end at $M \lsim -20$, where the measurements depart from
the Schechter function. The latter is due to residual contamination
from QSO light that the SED fitting could not eliminate completely.
The insets show the $1\sigma$, $2\sigma$ and $3\sigma$ confidence
regions on the $M^* - \alpha$ plane. As seen in Figure~\ref{fig:lfz},
there is a modest evolution in $M^*$ with redshift in both bands but
the leverage is not enough to constrain $\alpha$ to high
accuracy at higher redshifts. In all cases, the slope is consistent
with $\alpha = -1.1$.  Table~\ref{tbl:lf} lists the Schechter
parameters.

\begin{figure*}[]
\epsscale{1.05}
\plottwo{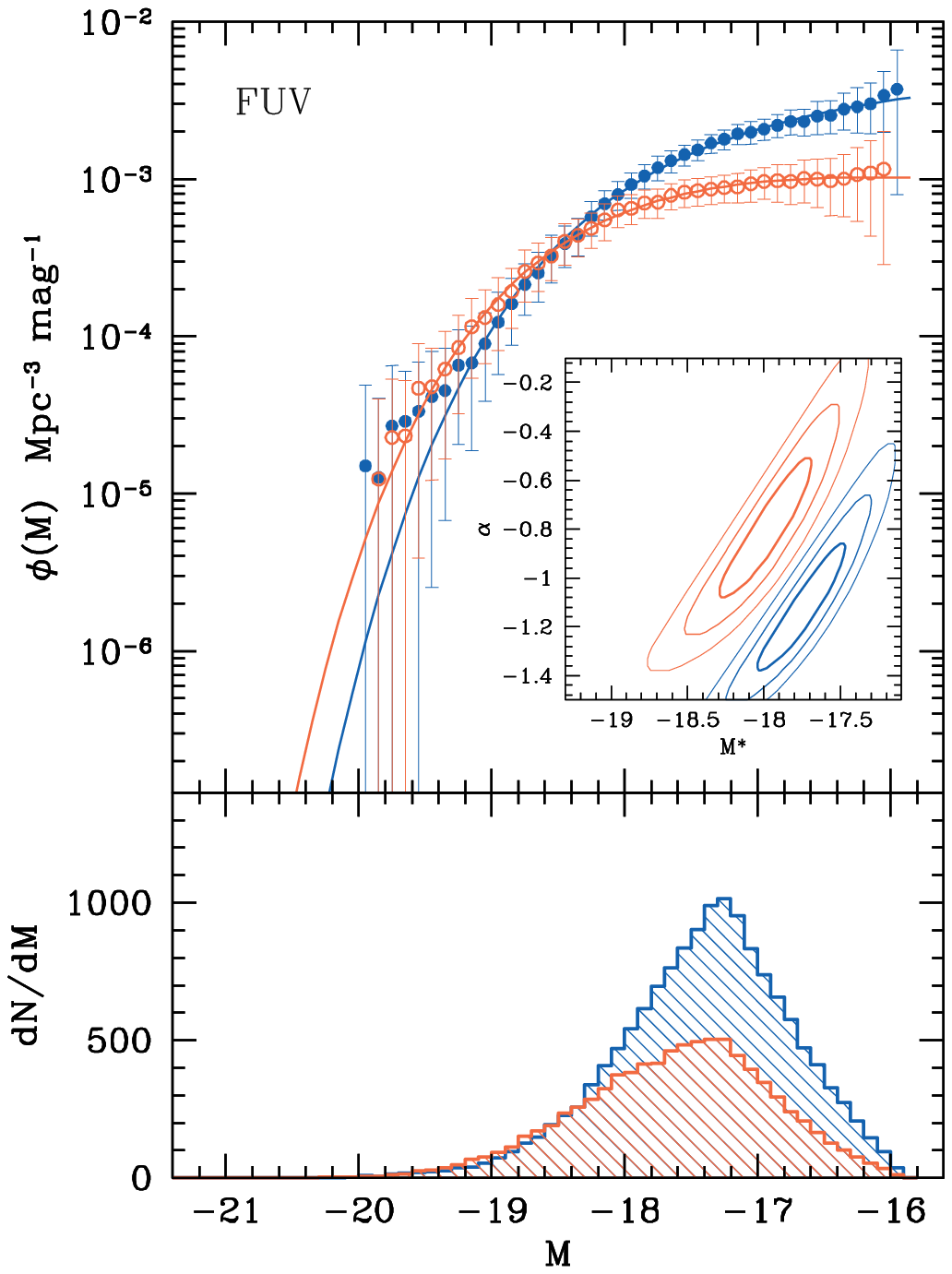}{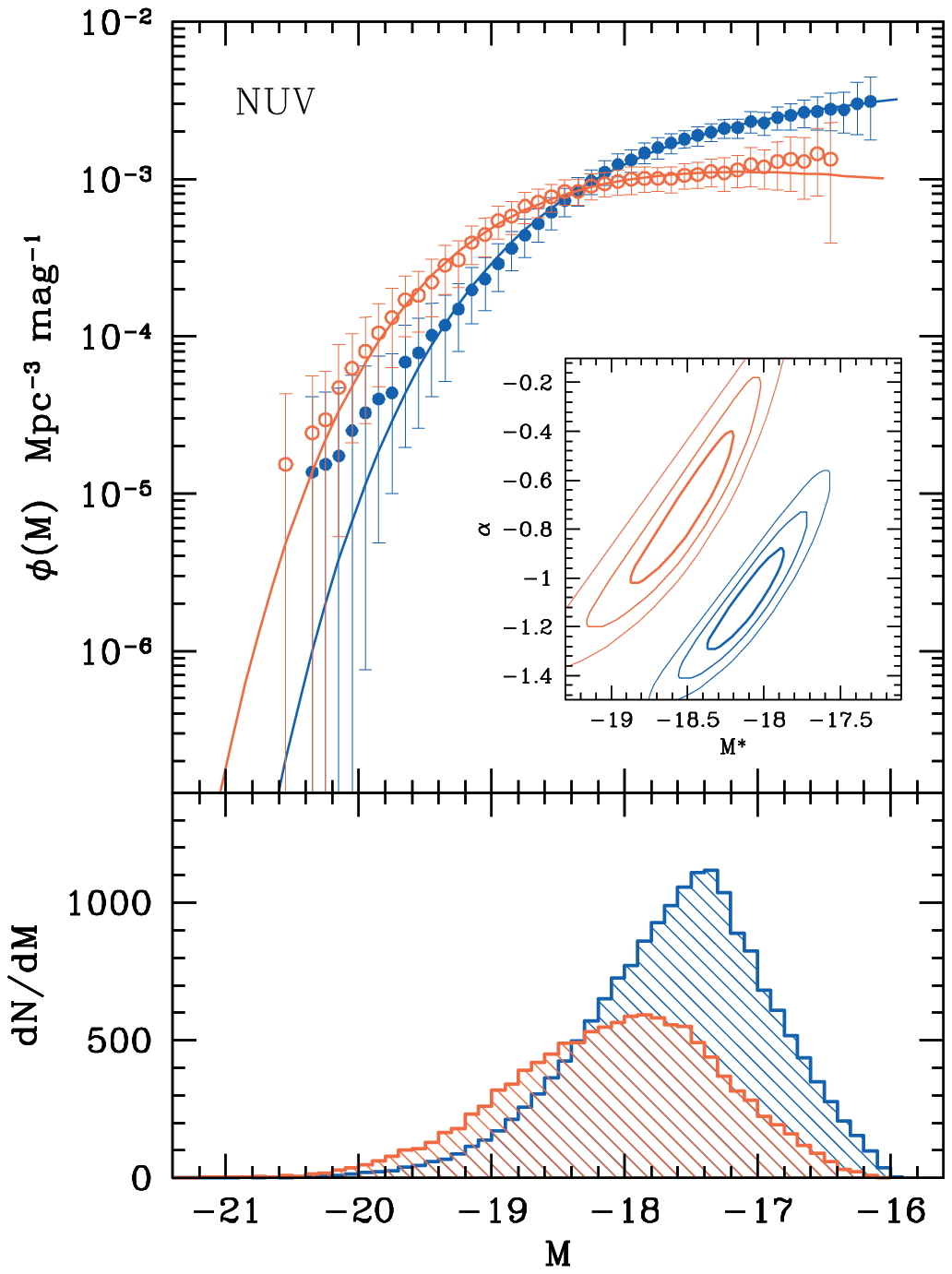}
\caption{The luminosity functions differ in both passbands for the
early and late type galaxies in the lowest redshift shell. The red
galaxy population is brighter and has a shallower faint-end slope
than the blue.}
\label{fig:lftype}
\end{figure*}

We further divide the lowest redshift \nuvmag{} and \fuvmag{} limited
samples into two spectral classes based on the assigned SEDs. The
early type galaxy class consists of objects with the five reddest
templates and late type galaxies with the five bluer SEDs, which
corresponds to a restframe color cut of $(u'\!-\!r')_0 =
1.7$. This technique is expected to be more robust than the actual
$(u'\!-\!r')_0$ discriminator as all multicolor information is used.
Figure~\ref{fig:lftype} shows the \fuvmag{} and \nuvmag{} luminosity
function for the early and late type galaxies.  We find that $M^*$ is
brighter for the red population by approximately 0.2 magnitudes in
\fuvmag{} and by 0.4 in \nuvmag{} and $\alpha$ is shallower than for
the blue population by 0.3, see Table~\ref{tbl:lf}. The marginalized
errors on both $M^*$ and $\alpha$ are large but the difference between
the joint distribution of parameters is significant (see insets of
Figure~\ref{fig:lftype}.)
When using the restframe colors to split galaxies into red and
blue, the plots exhibit the same features; the LFs at the
faint end ($M \gsim -18$) are essentially indistinguishable and the bright
end is also consistent with the template based results although the
difference is slightly less pronounced as expected due to the larger scatter.
The inverse concentration indices, simply the ratios of the radii
containing 50\% and 90\% of the Petrosian $r'$ fluxes, scatter
significantly but on average they are larger for galaxies in the blue
class than the red ones, $C^{-1}_{\rm{}blue} = 0.46$ and
$C^{-1}_{\rm{}red} = 0.42$. The red value is noticeably higher than
that of the elliptical SDSS galaxies, $C^{-1}_{} \approx{} 0.35$
\citep{strateva01}, which indicates that our redder class too contains
spiral galaxies.

\subsection{Luminosity Density}

\begin{figure}[b]
\epsscale{0.9}
\plotone{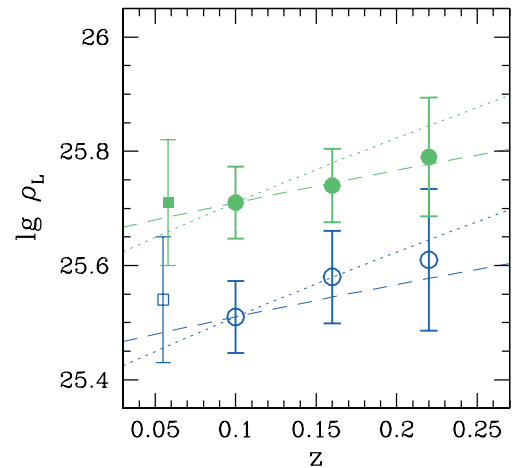}
\caption{The luminosity density as a function of redshift in \fuvmag{}
{\it{}(blue open circles)} and \nuvmag{} {\it{}(green filled circles)}
along with local GALEX results {\it{}(squares)} by \citet{wyder04}.
The dotted and dashed lines correspond to $(1 + z)^3$ and
$(1 + z)^{1.5}$, respectively, scaled to fit the lowest redshift
bin of this study.
}
\label{fig:zevol}
\end{figure}

We derive the mean luminosity density (LD) by integrating the
luminosity with the Schechter function, $\rho_L = \int \phi(L)\,L\,dL
= \phi^*\,L^*\,\Gamma(\alpha\!+\!2).$ In fact, the integral is
calculated for not just the optimal fit but over the whole range of
parameters. Weighting the results by the probability, $w =
\exp(-\chi^2/2)$, is essentially the same as using Monte-Carlo
realizations for estimating the errors, $\delta\rho_L^2 = \langle
\rho_L^2 \rangle_w - \langle \rho_L \rangle_w^2$. Since the redshift
range is limited, the errorbars on the higher redshift bins are rather
large. Table~\ref{tbl:lf} shows the LD measurements along with the
statistical errors and estimates for the systematics due to cosmic
variance that were derived similarly as in \citet{wyder04}. As seen in
Figure~\ref{fig:zevol}, both $\rho_{FUV}$ and $\rho_{NUV}$ increase
with redshift and are consistent with $(1+z)^3$ as well as
$(1+z)^{1.5}$.  The errorbars in the figure are the combinations of
the two sources of errors added in quadrature. They do not include
errors from calibration uncertainties of $\sim\!\!10\%$ that may
account for $\delta \lg \rho_L = 0.04$ in both bands.

\section{Discussion}

Using 2dF redshifts, the local GALEX studies by \citet{wyder04} and
\citet{treyer04} derived very consitent results with the present
findings. In the corresponding redshift ranges, the Schechter fits are
mostly within the 68\% confidence regions with perhaps the one
exception of $M^*$ in \nuvmag{}, which seems to be brighter based on
this photometric sample. The reason is that the different magnitude
cuts in the UV yield slightly different galaxy populations and our
sample has more redder galaxies, which makes $M^*$ brighter.

The LD measurements are also in good agreement with \citet{wyder04},
%see Figure~\ref{fig:zevol}, 
and the observed trend is consistent with
the GALEX Deep Imaging Survey results by \citet{arnouts04} and
\citet{schiminovich04} probing the higher redshift universe out to $z
= 1.2$. The evolution, in turn, is also consistent with results by
\citet{wilson02}, who find it to be proportional to $(1 +
z)^{1.7\pm{}1.0}$ for galaxies out to $z = 1.5$ at restframe
2500\AA{}.

Going from the observed UV luminosity function to a SFR function is
complicated by dust and the fact that the \fuvmag{} and \nuvmag{} light trace
stars forming on different timescales. We will address this problem in
subsequent GALEX papers.

\acknowledgments 

GALEX (Galaxy Evolution Explorer) is a NASA Small Explorer, launched
in April 2003. We gratefully acknowledge NASA's support for
construction, operation, and science analysis for the GALEX mission,
developed in cooperation with the Centre National d'Etudes Spatiales
of France and the Korean Ministry of Science and Technology.

\begin{deluxetable}{lllcccc}
%\tablefontsize{\footnotesize}
\tablecolumns{7}
\tablecaption{Schechter parameters and luminosity density}
\tablehead{
\colhead{Passband} & 
\colhead{Redshift} &
\colhead{Type} &
\colhead{$M^*$} &
\colhead{$\alpha$} & 
\colhead{$\lg \phi^*$ $[{\rm{}Mpc}^{-3}]$} & 
\colhead{$\lg \rho_L$ $[\frac{{\rm{}ergs}}{{\rm{}s\,Hz\,Mpc}^{3}}]$}
}
\startdata
\fuvmag{}...... & 0.07--0.13 & all & $-17.97\pm{}0.14$ & $-1.10\pm{}0.12$ & $-2.35\pm{}0.07$ & $25.51\pm{}0.02\pm{}0.06$ \\
& 0.13--0.19 & ........ & $-18.07\pm{}0.17$ & $-1.09\pm{}0.23$ & $-2.33\pm{}0.07$ & $25.58\pm{}0.07\pm{}0.04$ \\
& 0.19--0.25 & ........ & $-18.15\pm{}0.17$ & $-1.03\pm{}0.34$ & $-2.35\pm{}0.06$ & $25.61\pm{}0.12\pm{}0.03$ \\
\cline{2-7}
& 0.07--0.13 & early & $-17.98\pm{}0.20$ & $-0.80\pm{}0.19$ & $-2.74\pm{}0.08$ & $25.06\pm{}0.02\pm{}0.06$ \\
& 0.07--0.13 & late & $-17.74\pm{}0.19$ & $-1.12\pm{}0.17$ & $-2.47\pm{}0.09$ & $25.31\pm{}0.04\pm{}0.06$ \\
\\
\nuvmag{}...... & 0.07--0.13 & all & $-18.54\pm{}0.15$ & $-1.12\pm{}0.10$ & $-2.38\pm{}0.07$ & $25.71\pm{}0.02\pm{}0.06$ \\
& 0.13--0.19 & ........ & $-18.57\pm{}0.17$ & $-0.97\pm{}0.20$ & $-2.33\pm{}0.07$ & $25.74\pm{}0.05\pm{}0.04$ \\
& 0.19--0.25 & ........ & $-18.74\pm{}0.23$ & $-0.99\pm{}0.35$ & $-2.39\pm{}0.08$ & $25.79\pm{}0.10\pm{}0.03$ \\
\cline{2-7}
& 0.07--0.13 & early & $-18.53\pm{}0.23$ & $-0.73\pm{}0.21$ & $-2.66\pm{}0.08$ & $25.35\pm{}0.03\pm{}0.06$ \\
& 0.07--0.13 & late & $-18.11\pm{}0.17$ & $-1.09\pm{}0.14$ & $-2.48\pm{}0.08$ & $25.43\pm{}0.03\pm{}0.06$ \\
\enddata

\label{tbl:lf}
\end{deluxetable}

\end{document}